\documentclass[aps,twocolumn,nofootinbib]{revtex4}

\usepackage{graphicx}
\usepackage{dcolumn}
\usepackage{bm}

\usepackage{amssymb,amsfonts}
\usepackage{epsfig}
\usepackage{epstopdf}
\usepackage[all]{xy}
\usepackage{amsthm}
\usepackage{dcolumn}
\usepackage{hyperref}
\usepackage{url}
\usepackage{dsfont}

\newcommand{\be}{\begin{equation}}
\newcommand{\ee}{\end{equation}}
\newcommand{\bea}{\begin{eqnarray}}
\newcommand{\nn}{\nonumber}
\newcommand{\eea}{\end{eqnarray}}

\def\inbar{\,\vrule height1.5ex width.4pt depth0pt}
\def\IR{\relax{\rm I\kern-.18em R}}
\def\IC{\relax\hbox{$\inbar\kern-.3em{\rm C}$}}

\begin{document}

\title{``Massless" spin-2 field in de Sitter space}

\author{Hamed Pejhan$^{1}$} \author{Kazuharu Bamba$^2$\footnote{bamba@sss.fukushima-u.ac.jp}} \author{Surena Rahbardehghan$^3$\footnote{sur.rahbardehghan.yrec@iauctb.ac.ir}} \author{Mohammad Enayati$^3$}

\affiliation{$^1$Institute for Advanced Physics and Mathematics, Zhejiang University of Technology, Hangzhou 310032, China}

\affiliation{$^2$Division of Human Support System, Faculty of Symbiotic Systems Science, Fukushima University, Fukushima 960-1296, Japan}

\affiliation{$^3$Department of Physics, Science and Research Branch, Azad University, Tehran 1477893855, Iran}

\begin{abstract}
In this paper, admitting a de Sitter (dS)-invariant vacuum in an indefinite inner product space, we present a Gupta-Bleuler type setting for causal and full dS-covariant quantization of free ``massless" spin-2 field in dS spacetime. The term ``massless" stands for the fact that the field displays gauge and conformal invariance properties. In this construction, the field is defined rigorously as an operator-valued distribution. It is covariant in the usual strong sense: $\underline{U}_g \underline{{\cal{K}}} (X) \underline{U}_g^{-1} =  \underline{{\cal{K}}} (g.X)$, for any $g$ in the dS group, where $\underline{U}$ is associated with the indecomposable representations of the dS group, $SO_0(1,4)$, on the space of states. The theory, therefore, does not suffer from infrared divergences. Despite the appearance of negative norm states in the theory, the energy operator is positive in all physical states and vanishes in the vacuum.
\end{abstract}
\maketitle

\section{Introduction}
The subject of quantum field theory (QFT) in de Sitter spacetime is of paramount importance to the understanding of the early Universe as well as its present accelerated expansion (interpreted as the existence of a positive cosmological constant or a dark energy\footnote{For reviews on the so-called dark energy, see, \emph{e.g.,} \cite{Nojiri:2010wj,Nojiri:2006ri,Book-Capozziello-Faraoni,Capozziello:2011et,delaCruzDombriz:2012xy,Bamba:2012cp,Joyce:2014kja,Koyama:2015vza,Bamba:2015uma,Nojiri:2017ncd}.}). One of the most striking challenges of dS QFT is to formulate a completely satisfactory theory for ``massless" particles. Indeed, it is now quite universally believed that all physical theories that pretend to be fundamental make use of masslessness in one form or another; as massless photons and gravitons are basic to electrodynamics and to gravitation. On this basis, we are motivated to study ``massless" spin-2 particles in dS space.

The content of this paper is group theoretical. Considering the dS massless spin-2 field, it attempts to continue ``\emph{\`{a} la } Wigner" program for $SO_0(1,4)$. We begin our study by presenting the field equation as an eigenvalue equation of the coordinate-independent Casimir operators of the dS group (see section II). The Casimir operators carry the group-theoretical content of the theory. More technically, they enable us to classify the unitary irreducible representations (UIRs) of the dS group \cite{Dixmier,Takahashi289} based on two parameters $p$ and $q$ which, respecting the nature of the considered group representation, behave like a spin (s) and a mass (m) in the Minkowskian limit. This group-theoretical structure is not coordinate dependent. However, in order to make the structure explicit, we shall utilize the dS ambient space coordinates. Interestingly, this approach allows us to clarify what is meant by the concept of masslessness in dS space. This concept can be realized by examining several criteria: conformal extension, Poincar$\acute{e}$ contraction, light-cone propagation, and gauge invariance. The latter is intimately related to the dS indecomposable representations. A comprehensive discussion on the masslessness criteria can be found in \cite{angelo,Flato415}.

Massless particles in Minkowski space are associated with UIRs of the Poincar$\acute{e}$ group, with zero mass and with discrete helicity (the Poincar$\acute{e}$ massless representations). These representations are the only ones of the Poincar$\acute{e}$ group that have unique extensions to the conformal group. More exactly, it is proved that any system that is invariant under a massless representation of the Poincar$\acute{e}$ group is invariant under a uniquely determined UIR of the conformal group \cite{angeloflato}. In this sense, the only representations of the dS group that hold this property, in the Dixmier's notation \cite{Dixmier}, are $\Pi_{p,q}^{+}\oplus \Pi_{p,q}^{-}$ when $q=p=s$ (in our case $q=p=2$).\footnote{Note that, the signs $+$ and $-$ stands for the two types of helicity.} These representations are associated with the discrete series representations of the dS group. Moreover, they contract smoothly to the Poincar$\acute{e}$ massless representations in the limit of vanishing curvature \cite{levy,barut}. From now on, we refer to them as the dS massless representations.

It should be noted that the physical representations are unitary and as already pointed out they belong to the dS massless representations, however, this does not mean that our theory is conformally invariant. The point is in fact the gauge invariance property of the theory (another feature of masslessness that make sense in dS space). Let us be more precise. If the dS massless spin-2 representation is realized in terms of the traceless-transverse rank-2 tensor field, the solutions to the associated wave equation result in a singularity due to the divergencelessness condition. This condition is required to relate the tensor field to the dS massless representations. To fix this problem, the divergencelessness condition must be ignored. The modified field equation then becomes gauge invariant, and one is free to use a gauge-fixing parameter $c$. Note that, the gauge-invariant subspace makes an indecomposable structure unavoidable.

On this basis, the theory admits three spaces of solutions: the space of the gauge and the $c$-independent divergencelessness solutions, respectively, denoted by $V^g$ and $V^{ci}$, and the space of the $c$-dependent solutions which are not divergenceless $V^{cd}$, so that, $V^g \subset V^{ci} \subset V^{cd}$. The gauge solutions are orthogonal to the whole divergencelessness solutions including themselves. The solutions associated with each part are explicitly calculated in section III. It is discussed that the gauge-fixing parameter $c=2/5$, according to the group representation theory, leads to the minimal covariant structure of the space of solutions. Any departure from this choice results in logarithmic states which imply reverberation inside the light cone. Of course, whatever $c$ one chooses, the propagation of the divergencelessness solutions is confined to the light cone.

In section IV, the role of dS invariance is considered in detail. We show that, considering the standard positive frequency solutions (with respect to the conformal time), the space of solutions is not invariant under the action of the dS group $SO_0(1,4)$; all negative frequency solutions are unavoidably generated. This difficulty, known as the ``zero-mode" problem, is indeed inherited from the dS minimally coupled scalar (MCS) field, which is appeared as a structure function in the space of the dS massless spin-2 field solutions. To circumvent this problem, respecting the minimal requirements for a canonical quantization, we present a Gupta-Bleuler type formalism based on weaker conditions, which does not prohibit negative frequency solutions in the theory. This structure is simply called the Krein-Gupta-Bleuler (KGB) structure. Here, the term ``Krein", more exactly the Krein space, stands for direct sum of the Hilbert and the anti-Hilbert spaces associated with the dS massless spin-2 field. On this larger framework, the Krein space, each of three spaces of solutions ($V^g$, $V^{ci}$ and $V^{cd}$) would be invariant under the action of the dS group. This action is indecomposable; that is, there is no invariant subspace that is complementary to $V^g$ in $V^{ci}$ or to $V^{ci}$ in $V^{cd}$. To ensure a reasonable interpretation of the theory, now, we need to specify the subspace of physical modes. Demanding the positivity requirement and also being invariant under the action of the dS group as well as the gauge transformation, this space is given in section IV. We show that the central part $V^{ci}/V^g$ contains all the physical modes (of course, it is not restricted to them).

In section V, the Fock space structure and also the field operator are constructed. The field fulfills the conditions of: a) locality, b) covariance, c) transversality and d) tracelessness. It is therefore free of infrared divergences. Again, in our KGB quantization scheme the positivity requirement in the definition of the field has been ignored. In this regard, it must be underlined that the field itself is not observable (it is gauge dependent). The stress tensor however is. We discuss this matter in section VI and show that the KGB quantization scheme provides an automatic and covariant renormalization of the stress tensor, so that, the vacuum energy of the free field vanishes without any reordering nor regularization, and on the physical states it is always positive. This assures a reasonable physical interpretation of the theory. We finally discuss our result in section VII.

\section{Group Content of de Sitterian Relativity}
The dS space is conveniently seen as (the covering space of) a one-sheeted hyperboloid embedded in a five-dimensional Minkowski space
$$ {M_{H}}  = \{ x \in \mathbb{R}^5 ; x^2={\eta}_{\alpha\beta} {x^{\alpha}} {x^\beta} =  -H^{-2}\},$$
where $H$ stands for the Hubble constant, $\alpha ,\beta =0,1,2,3,4,$ and $\eta_{\alpha\beta}=$ diag$(1,-1,-1,-1,-1)$. The induced metric reads
\begin{equation}\label{2.2}
ds^2=\eta_{\alpha\beta}dx^{\alpha}dx^{\beta}|_{x^2=-H^{-2}}={g}_{\mu\nu}dX^{\mu}dX^{\nu},
\end{equation}
where the $X^\mu$'s are intrinsic spacetime coordinates ($\mu,\nu=0,1,2,3$).

This description of the dS spacetime, a pseudo-sphere in the bulk (a higher-dimensional Minkowski space), is called the ambient space notations. In this approach, a tensor field ${\cal K}(x)$ (dS field) is considered as a homogeneous function of the $\mathbb{R}^5$-variables $x^\alpha$ as follows
\begin{equation} \label{homo}
x^\alpha \frac{\partial}{\partial x^\alpha}{\cal{K}}(x) = x\cdot\partial {\cal{K}}(x) = \varrho{\cal{K}}(x),
\end{equation}
in which $\varrho$ is an arbitrarily selected degree. For the sake of simplicity, we select $\varrho = 0$, for which, the d'Alembertian operator $\square\equiv \nabla_\mu \nabla^\mu$ on dS intrinsic spacetime ($\nabla_\mu$ being the covariant derivative) corresponds to its counterpart $\square_5 \equiv \partial^2$ on $\mathbb{R}^5$. In addition, the dS fields must satisfy the transversality condition $x.{\cal K}(x)=0$, to ensure that the direction of them lies in the dS tangent space. Because of the importance of this transversality, the symmetric and transverse projector $\theta_{\alpha\beta}=\eta_{\alpha\beta}+H^2x_\alpha x_\beta$ is defined to permit one to construct transverse entities like the transverse derivative on dS space, $\bar{\partial}_\alpha =\theta_{\alpha\beta}\partial ^\beta =\partial _\alpha +H^2x_\alpha x\cdot \partial$; $x \cdot \bar{\partial} = 0$. Note that, $\theta_{\alpha\beta}$ is the only tensor which is linked to the dS metric, so that, ${g}_{\mu\nu} = x_{\mu}^{\alpha}x_{\nu}^{\beta}\theta_{\alpha\beta}$ with $x_{\mu}^{\alpha}=\partial x^{\alpha}/\partial X^{\mu}$. Similarly, any intrinsic tensor field ${\cal{K}}_{\mu_1...\mu_r}(X)$ can be locally characterized by the transverse tensor field ${\cal{K}}_{\alpha_1...\alpha_r}(x(X))$ through the following relation,
\begin{equation} \label{666}
{\cal{K}}_{\mu_1...\mu_r}(X)= x_{\mu_1}^{\alpha_1}...x_{\mu_r}^{\alpha_r} {\cal{K}}_{\alpha_1...\alpha_r}(x(X)).
\end{equation}

As we pointed out earlier, the aim of this section is to present the dS massless spin-2 (the traceless and symmetric rank-2 massless tensor) field equation as an eigenvalue equation of the dS Casimir operator. The dS relativity group, $O(1,4)$, is the ten-parameter group with two Casimir operators. In this paper, we will only consider its connected component $SO_0(1,4)$ and focus on its quadratic (or second order) Casimir operator denoted by $Q_r$. The index `$r$' herein means the carrier space is constituted by rank-$r$ tensors. This Casimir operator on the ambient space reads \cite{Gazeau2533,Gazeau507}
\begin{eqnarray}\label{Casimirope}
Q_r \equiv -\frac{1}{2}L_{\alpha\beta}^{(r)}L^{{(r)}{\alpha\beta}},
\end{eqnarray}
where the ten infinitesimal generators $L_{\alpha\beta}^{(r)}=M_{\alpha\beta}+S_{\alpha\beta}^{(r)}$ are the self-adjoint representatives of the killing vectors. The orbital part is given by
\begin{equation} \label{M}
M_{\alpha\beta}=-i(x_\alpha \partial _\beta -x_\beta \partial _\alpha),
\end{equation}
and the spinorial part $S_{\alpha\beta}^{(r)}$ acts on the tensor indices as follows
\begin{eqnarray}\label{self-ad}
S_{\alpha\beta}^{(r)}{\cal K}_{\alpha _1 ... \alpha _r}=-i\displaystyle\sum_{i=1}^{r}(\eta_{\alpha\alpha _i}{\cal K}_{\alpha _1...(\alpha _i\rightarrow \beta)...\alpha_r}\nn\\
-\eta_{\beta\alpha_i}{\cal K}_{\alpha _1...(\alpha _i\rightarrow \alpha)...\alpha_r}).
\end{eqnarray}

The Casimir operator commutes with all generators of the dS group and, as a consequence, it has a constant value on all the states in each UIR. Hence, the eigenvalues of $Q_r$ can be considered to classify the UIR's. More  precisely, the states of the dS UIRs lie among the solutions of the following dS-invariant equation \cite{Gazeau329,Gazeau2533}
\begin{eqnarray}\label{CasimireqK}
(Q_r-\langle Q_r\rangle){\cal K}=0,
\end{eqnarray}
supplemented with the divergencelessness condition ($\partial\cdot {\cal K}=0$). Note that, this condition along with transversality of dS fields imply the tracelessness condition  \cite{Gazeau329}
$${\cal K}^{\prime}_{\alpha_1 ...\alpha_{r-2}}\equiv \eta^{\alpha_{r-1}\alpha_r}{\cal K}_{\alpha_1...\alpha_{r-2}\alpha_{r-1}\alpha_r}=0.$$
Following Dixmier \cite{Dixmier} the UIR's then can be labelled by a pair of parameters $p$ and $q$, in terms of the eigenvalues of $Q_r$,
\begin{eqnarray}\label{3}
\langle Q_r\rangle=[-p(p+1)-(q+1)(q-2)],
\end{eqnarray}
with $2p \in\mathbb N$ and $q\in\mathbb C$. According to the possible values of the parameters $p$ and $q$, the dS UIRs can be split into three types of inequivalent categories, namely, the principal, complementary and discrete series. For the principal and complementary series, the $H = 0$ contraction limit (vanishing curvature limit) compels the value of $p$ to bear the meaning of spin. In the case of the discrete series, however, label $q$ has a spin meaning. One can get a detailed discussion about the mathematical and physical principles underlying the contraction between the dS and Poincar$\acute{e}$ group in Refs. \cite{levy} and \cite{bacry}.

The spin-2 tensor representations associated with our study in this paper are as follows:

I) The UIR's $U^{2,\nu}$ in the principal series, with $p = s = 2$ and $q = \frac{1}{2} + i\nu$, correspond to
\begin{equation} \langle Q_2^{\nu}\rangle = {\nu}^2 - \frac{15}{4}, \;\;\; {\nu}\in \mathbb{R}. \end{equation}
Note that, $U^{2,\nu}$ and $U^{2,-\nu}$ are equivalent.

II) The UIR's $V^{2,q}$ in the complementary series, with $p = s = 2$ and $ q - q^2 = \mu $, correspond to
\begin{equation} \langle Q_2^{\mu} \rangle = q - q^2 -4 \equiv{ {\mu} -4}, \;\;\; 0 < \mu < \frac{1}{4}. \end{equation}

III) The UIR's $\Pi^{\pm}_{p,2}$ in the discrete series, with $q = s = 2$, correspond to
\begin{equation} \label{2.12} \langle Q^p_2\rangle = -p(p+1).\end{equation}

Based on the contraction of the group representations, the following ``mass" formula has been proposed by Garidi \cite{What} in terms of the dS UIR parameters $p$ and $q$:
\begin{equation} \label{mass}
m_H^2= \langle Q_r \rangle - \langle Q_r^{p=q}\rangle = [(p-q)(p+q-1)]\hbar^2H^2/c^4.
\end{equation}

Since we have set the zero of the mass parameter $m_H$ according to the lowest value of the Casimir operator, \emph{i.e.,} for $p = q$ which corresponds to the conformal massless case, we are insured that every dS UIRs which are meaningful from a Minkowskian viewpoint are labelled by $m_H^2 \geq0$. On this basis, the massless spin-2 field in dS space corresponds to the representation $\Pi^\pm_{2,2}$ with $\langle Q_2\rangle =-6$ (from now on, we simplify our notations by considering $ Q_2^{p=2}\equiv Q_2$). The dS representation $\Pi^+_{2,2}$ has indeed a unique extension to a direct sum of two UIRs of the conformal group, namely ${\cal C}^>(3,2,0)$ and ${\cal C}^<(-3,2,0)$, respectively, associated with positive and negative energies.\footnote{The compact subgroup of the conformal group $SO(2,4)$ is determined by $SO(2)\otimes SO(4)$. Considering $E$ as the eigenvalues of the conformal energy generator of $SO(2)$ and $(j_1,j_2)$ as the $(2j_1+1)(2j_2+1)$ dimensional representation of $SO(4)=SU(2)\otimes SU(2)$, the symbols ${\cal C}(E,j_1,j_2)$ stand for irreducible projective representation of $SO(2,4)$.} That extension is equivalent to the conformal extension of a massless UIR of the Poincar\'e group with helicity $+ 2$, symbolized by ${\cal P}^>(0, 2)$ and ${\cal P}^<(0,2)$. Symbols ${\cal P}^{ \stackrel{>} {<}}(0,\pm2)$ denote the Poincare massless representations with helicity $\pm2$ and with positive (respectively negative) energy. Similar arguments can be applied to the representation $\Pi^-_{2,2}$. The following diagrams present these correspondences  \cite{barut,levy}
\begin{equation}
\left.
\begin{array}{ccccccc}
&& {\cal C}^>(3,2,0)& &{\cal C}^>(3,2,0)&\hookleftarrow &{\cal P}^{>}(0,2)\\
\Pi^+_{2,2} &\hookrightarrow &\oplus&\stackrel{H=0}{\longrightarrow} & \oplus  & &\oplus\\
&& {\cal C}^<(-3,2,0)& & {\cal C}^<(-3,2,0) &\hookleftarrow &{\cal
P}^{<}(0,2),\\
\end{array}
\right.
\end{equation}

\begin{equation}
\left.
\begin{array}{ccccccc}
&& {\cal C}^>(3,0,2)& &{\cal C}^>(3,0,2)&\hookleftarrow &{\cal P}^{>}(0,-2)\\
\Pi^-_{2,2} &\hookrightarrow &\oplus&\stackrel{H=0}{\longrightarrow}&\oplus &&\oplus\\
&& {\cal C}^<(-3,0,2)&& {\cal C}^<(-3,0,2)&\hookleftarrow &{\cal P}^{<}(0,-2),\\
\end{array}
\right.
\end{equation}
the arrows $\hookrightarrow $ determine unique extension.

Here, it must be underlined that Eq. (\ref{CasimireqK}) is not suitable for the massless spin-2 field. In fact, for the Casimir operator eigenvalue $\langle Q_2\rangle =-6$, the solution to (\ref{CasimireqK}), \emph{i.e.},
\begin{eqnarray}\label{CasimireqKsol}
{\cal K}_{\alpha\beta}(x)={\varepsilon}_{\alpha\beta}^1(x,Z,\xi)\Phi(x) \hspace{3cm} \nn\\
+\frac{1}{(\langle Q_2\rangle +6)}\varepsilon _{\alpha\beta}^2(x,Z,\xi ,\bar\partial)\Phi(x),
\end{eqnarray}
results in a singularity due to the term $1/(\langle Q_2\rangle +6)$ \cite{Garidi3838}. Here $\varepsilon _{\alpha\beta}^1$ and $\varepsilon _{\alpha\beta}^2$ are operators that act on the scalar field $\Phi (x)$. $Z$ is a constant vector in ambient space and $\xi \in \mathbb{R}^5$ lies on the null cone $\xi ^2=0$. Note that, $\varepsilon ^1$, $\varepsilon ^2$ and $\Phi (x)$ also contain the parameters $p$ and $q$, but which do not diverge for $p=q=2$ \cite{Garidi3838}. This means that the subspace supplemented with the condition $\partial\cdot {\cal K} =0$ is not sufficient for the construction of a quantum massless spin-2 field. To fix this problem, we need to drop the divergencelessness condition, \emph{i.e.}, $\partial\cdot {\cal K}\neq 0$ \cite{Garidi3838}. Hence, the modified equation of (\ref{CasimireqK}) takes the form
\begin{eqnarray}\label{Fieq}
(Q_2+6){\cal K}+D_2{\partial}_2\cdot {\cal K}=0,
\end{eqnarray}
where the generalized gradient on the dS hyperboloid is $D_2\equiv {\cal S}(D_1-x)$, in which the operator ${\cal S}$ is the symetrizer (${\cal S}\xi _\alpha \omega _\beta =\xi _\alpha \omega _\beta +\xi_\beta \omega _\alpha $) and $D_1=H^{-2}\bar{\partial}$. The action of the generalized divergence on a general rank-2 tensor field is ${\partial}_2\cdot {\cal K}=\partial\cdot {\cal K}-H^2x{\cal K}^\prime-\frac{1}{2}\bar{\partial}{\cal K}^\prime $.

By using the following identities
\begin{eqnarray}\label{auxi1}
{\partial}_2\cdot D_2\Lambda^g=-(Q_1+6)\Lambda^g, \;\; Q_2D_2\Lambda^g=D_2Q_1\Lambda^g,
\end{eqnarray}
one can simply show that Eq. (\ref{Fieq}) is invariant under the general gauge transformation ${\cal K}\rightarrow {\cal K}+D_2\Lambda^g$, in which $\Lambda^g$ is an arbitrary vector field. It is known that, because of this gauge symmetry the canonical quantization of the massless spin-2 field becomes impossible. Eq. (\ref{Fieq}), therefore, have to be modified in order to circumvent this problem as follows
\begin{eqnarray}\label{FE}
(Q_2+6){\cal K}^{cd}+cD_2{\partial}_2\cdot {\cal K}^{cd}=0,
\end{eqnarray}
where $c$ is a constant called the ``gauge-fixing parameter" and added to the theory to restrict the subspace of gauge solutions. We denote by ${\cal K}^{cd}$ the general solutions to (\ref{FE}) to remind us that they are $c$-dependent. The important point to note here is that Eq. (\ref{FE}) is exactly the ambient counterpart of the transverse-traceless sector of the dS linearized Einstein equation in the context of the most general gauge-fixing functionals \cite{I,II}.

It is obvious that the field equation (\ref{FE}) becomes fully gauge invariant if we put $c=1$. For other choices, the tensor field ${\cal K}^{cd}$ would be traceless \cite{Gazeau329} and associated with an indecomposable representation of the dS group; from now on, we regard $c\neq 1$. A structure analogous to that of the Gupta-Bleuler triplets of Minkowski QED or of dS QED then could appear in the space of solutions \cite{BinegarGUPTA,Gazeau1847}. We will show that, see section III, the general solutions to (\ref{FE}) still suffer from logarithmic singularities, and one can eliminate them by adopting a suitable choice of $c$.

\section{The Field Solution}
In the previous section, we showed that in the case of the dS massless spin-2 field, as expected for a massless field, one faces three kinds of solutions: the gauge solutions, the divergencelessness solutions, and the solutions which are not divergenceless. The latter is simply called the general solution. We here explicitly calculate each of these three kinds of solutions (again, for $c\neq 1$).

The general solution to the dS massless spin-2 field equation (\ref{FE}) can be written in terms of two tensors of rank-1 ($K$ and $K^g$) and a rank-0 tensor ($\phi_1$) through the following linearly independent formula \cite{Gazeau2533}
\begin{eqnarray}\label{recuform}
{\cal K}^{cd}=\theta\phi_1 +{\cal S}\bar{Z}K+D_2K^g,\;\; \partial_2\cdot {\cal{K}}^{cd}\neq0.
\end{eqnarray}
The operators $\theta$, ${\cal S}\bar{Z}$ and $D_2$ make a symmetric transverse rank-2 field from the scalar field and the vector fields, respectively. Here, $K^g$ and $K$ are transverse ($x\cdot K^g=0=x\cdot K$) and the vector field $K$ is divergenceless, \emph{i.e.}, $\partial\cdot K=0$.\footnote{Note that, $x\cdot K=0$ implies that $\partial\cdot K = \bar\partial\cdot K$.} Moreover, we have
\begin{eqnarray}\label{4.2}
2\phi_1 + Z\cdot K + H^{-2}\bar\partial\cdot K^g = 0.
\end{eqnarray}
It is a direct consequence of the tracelessness condition, $({\cal K}^{cd})' = 0$.

After putting (\ref{recuform}) into (\ref{FE}) and using (\ref{auxi1}) and the following relations
\begin{eqnarray}\label{help1}
Q_2\theta\phi_1=\theta Q_0\phi_1, \;\;\;\; \partial_2\cdot\theta\phi_1=-H^2D_1\phi_1,
\end{eqnarray}
\begin{eqnarray}\label{help2}
Q_2{\cal S}\bar{Z}K&=&{\cal S}\bar{Z}(Q_1-4)K -2H^2D_2x\cdot ZK+4\theta Z\cdot K,\hspace{0.5cm}\nn\\
\partial_2\cdot {\cal S}\bar{Z}K&=& TZ\cdot \bar\partial K - H^2 D_1 Z\cdot K + 5H^2x\cdot ZK,
\end{eqnarray}
one obtains
\begin{eqnarray}\label{s1}
(Q_0+6)\phi_1=-4Z\cdot K,
\end{eqnarray}
\begin{eqnarray}\label{s2}
(Q_1+2)K=0,
\end{eqnarray}
and
\begin{eqnarray}\label{s3}
(1-c) (Q_1+6)K^g= (2-5c) H^2x\cdot ZK \hspace{1.5cm}\nn\\
+ c (-\frac{1}{2}H^2D_1\phi_1-TZ\cdot\bar{\partial}K) + \Xi^g,
\end{eqnarray}
where $TZ\cdot\bar{\partial}K=Z\cdot\bar{\partial}K-H^2xZ\cdot K$, and $\Xi^g$ is an arbitrary vector field which is due to the canceling property of $D_2$,
\begin{eqnarray}\label{Xidisa}
D_2 \Xi^g =0.
\end{eqnarray}

The scalar field $\phi_1$ in Eq. (\ref{s1}) is completely determined by
\begin{eqnarray}\label{s1s}
\phi_1=-\frac{2}{3}Z\cdot K,
\end{eqnarray}
where we use this fact that Eq. (\ref{s2}) combined with $\bar\partial\cdot K=0$ imply $Q_0K=0$. Therefore, we have
\begin{eqnarray}\label{sls''}
Q_0\phi_1=-\frac{2}{3} Q_0 Z\cdot K=0.
\end{eqnarray}
Considering $K^g = \widetilde{K}^g + \Lambda^g,$ while $(1-c) (Q_1+6)\Lambda^g = \Xi^g$ ($x\cdot \Lambda^g=0$, $\bar\partial\cdot\Lambda^g=0$), we can rewrite the inhomogeneous equation (\ref{s3}) as
\begin{eqnarray}\label{inhomo}
(Q_1+6)\widetilde{K}^g=[1/(1-c)][-c,c/3,2-5c],
\end{eqnarray}
where $[a,b,e]\in E$; the space $E$ is the three-dimensional space generated by a linear combination of a set of three basic functions
$$[a,b,e]=aTZ\cdot\bar{\partial}K+bH^2D_1Z\cdot K+eH^2x\cdot ZK.$$
The space $E$ is invariant under the action of $(Q_1+6)$,
\begin{eqnarray}\label{inv1}
(Q_1+6)TZ\cdot\bar{\partial}K=[6,2,0],
\end{eqnarray}
\begin{eqnarray}\label{inv2}
(Q_1+6)H^2D_1Z\cdot K=[0,6,0],
\end{eqnarray}
\begin{eqnarray}\label{inv3}
(Q_1+6)H^2x\cdot ZK=[-2,0,0].
\end{eqnarray}
Therefore, the solution to Eq. (\ref{inhomo}), $\widetilde{K}^g=[v,u,w]$, is simply obtained with respect to the following system
\begin{eqnarray}\label{mat}
\left(\begin{array}{ccc} 6 &  0 & -2 \\ 2 & 6 & 0 \\ 0 & 0 & 0 \\
\end{array}\right) \left(\begin{array}{ccc} v \\ u \\ w \\ \end{array}\right)=\frac{1}{(1-c)}\left(\begin{array}{ccc} -c \\ c/3 \\ 2-5c \\ \end{array}\right)
\end{eqnarray}
Note that, the matrix determinant is zero. This implies that, Eq. (\ref{inhomo}) gives a solution inside $E$ as long as we adjust the gauge-fixing parameter to $c=2/5$ (the simplest structure). But, as we will see, it is interesting to study the general solution with an arbitrary value for $c$ (the general structure). In the following, we investigate both of them.

\subsection{The Simplest Structure; $c= 2/5$}
With the value $c=2/5$, the solution to Eq. (\ref{inhomo}) would be
\begin{eqnarray}\label{inhomosolution}
\widetilde{K}^g= [0 ,1/27 , 1/3]+ \kappa \Lambda^\circ,
\end{eqnarray}
where $\kappa$ is an arbitrary constant, and $\Lambda^\circ$ is a function inside $E$ that verifies
\begin{eqnarray}\label{lambda0}
(Q_1+6)\Lambda^\circ= 0.
\end{eqnarray}
This solution is given up to a multiplicative constant, as follows
\begin{eqnarray}\label{lambda00}
\Lambda^\circ=[1,-1/3,3].
\end{eqnarray}
It should be noted that, due to the appearance the arbitrary $\Lambda^g$ in $K^g$, the term $\kappa \Lambda^\circ$ can be dropped. We here, however, consider the most general case, which allows us to clarify the group theoretical meaning of $\widetilde{K}^{g}$. In this regard, we need to find the equation satisfied by $\widetilde{K}^{g}$. First of all, $\Lambda^\circ$ is divergenceless, $\widetilde{K}^{g}$ however is not,
\begin{eqnarray}\label{nnn}
\bar\partial\cdot\Lambda^\circ=0,\;\; D_1 \bar\partial\cdot\widetilde{K}^{g}= [0,1/3,0].
\end{eqnarray}
The latter is compatible with (\ref{4.2}) (when it is combined with (\ref{s1s})). On the other hand, we have
$$ L^{(2)}_{\alpha\beta} (H^2 D_2 D_1 Z\cdot K) = H^2 D_2 D_1 M_{\alpha\beta} Z\cdot K.$$
This equation reveals that, inside the solutions to (\ref{FE}), the term $H^2 D_2 D_1 Z\cdot K $ carries the same representation as $\phi_1$. More exactly, $H^2 D_2 D_1 Z\cdot K $ does not carry any spin; it is completely determined by its scalar content. We denote this scalar part of $\widetilde{K}^{g}$ by $\widetilde{K}^{1g}$ and call $\widetilde{K}^{2g}$ what is left from it; $\widetilde{K}^{1g} + \widetilde{K}^{2g} = \widetilde{K}^{g}$.

Now, let us make $\widetilde{K}^{2g}$ explicit. Applying $Q_1$ on $\widetilde{K}^{2g}$, with regard to Eq. (\ref{inhomosolution}), results in
\begin{eqnarray}\label{mmm}
Q_1 \widetilde{K}^{2g}= [-2/3-6\kappa,2\kappa,-2-18\kappa].
\end{eqnarray}
To calculate the above equation, relation (\ref{inv1}), (\ref{inv2}) and (\ref{inv3}) have been utilized. $\widetilde{K}^{2g}$ is not divergenceless, therefore, we have to combine (\ref{mmm}) with (\ref{nnn}) (note that, $\bar\partial\cdot \widetilde{K}^{1g}= \bar\partial\cdot (H^2 D_1 Z\cdot K) = - H^2 Q_0 Z\cdot K = 0$). In this sense, setting $\kappa=-1/9$, we obtain
\begin{eqnarray}\label{bbb}
Q_1 \widetilde{K}^{2g} + \frac{2}{3}D_1 \bar\partial\cdot\widetilde{K}^{2g} =0.
\end{eqnarray}
Now, the group theoretical meaning of $\widetilde{K}^{2g}$ is obvious. It carries the massless representations with spin-1 and gauge fixing parameter $c=2/3$. This value exactly corresponds to the minimal structure. See \cite{Garidi032501}, for detailed discussions.

Consequently, in the case of $c=2/5$ and $\kappa=-1/9$, the general solution to Eq. (\ref{s3}) would be $K^g = \widetilde{K}^{1g} + \widetilde{K}^{2g} + \Lambda^g$, in which
\begin{eqnarray}\label{K^1g}
\widetilde{K}^{1g} = [0,2/27,0],\;\; \widetilde{K}^{2g} = [-1/9,0,0].
\end{eqnarray}
On this basis, the general solution to the field equation (\ref{FE}), in the simplest case, can be written as
\begin{eqnarray} \label{ci;}
{\cal K}^{cd;c=\frac{2}{5}}= \theta\phi_1 +{\cal S}\bar{Z}K + \frac{2}{27} H^2 D_2D_1 Z\cdot K \hspace{1cm} \nn\\
 + D_2( \widetilde{K}^{2g} + \Lambda^g).
\end{eqnarray}
Note that, the gauge solutions ${\cal{K}}^g = D_2 \Lambda^g$ obey
\begin{eqnarray} \label{gaugeq}
D_2 (Q_1+6) \Lambda^g = 0.
\end{eqnarray}
For the sake of simplicity and with respect to our group theoretical approach, however, we choose
\begin{eqnarray}\label{GS}
(Q_1+6)\Lambda^g=0.
\end{eqnarray}

It is now interesting to identify the divergencelessness part of the solutions. Utilizing (\ref{auxi1}), (\ref{help1}) and (\ref{help2}), we have
\begin{eqnarray}\label{conphy}
\partial_2\cdot {\cal K}^{cd;c=\frac{2}{5}}=\frac{1}{1-\frac{2}{5}} \Big( TZ\cdot \bar\partial K -\frac{1}{3}H^2 D_1 Z\cdot K \;\;\;\;\nn\\
 + 3H^2x\cdot ZK \Big).
\end{eqnarray}
By comparing (\ref{conphy}) with (\ref{lambda00}), it is obvious that $\Lambda^\circ =(1-\frac{2}{5})\partial_2\cdot {\cal K}^{cd;c=\frac{2}{5}}$.

Now, combining (\ref{conphy}) and (\ref{ci;}), we have
\begin{eqnarray} \label{ci;''}
{\cal K}^{cd;c=\frac{2}{5}}&=& \Big(\theta\phi_1 +{\cal S}\bar{Z}K + \frac{1}{27} H^2 D_2D_1 Z\cdot K \nn\\
&& + \frac{1}{3}H^2 D_2 x\cdot ZK\Big) + \Big( \frac{\frac{2}{5}-1}{9} D_2\partial_2\cdot {\cal K}^{cd;c=\frac{2}{5}}\Big) \nn\\
&\equiv & {\cal K}^{ci} + \frac{\frac{2}{5}-1}{9} D_2\partial_2\cdot {\cal K}^{cd;c=\frac{2}{5}}.
\end{eqnarray}
The divergencelessness solutions ${\cal K}^{ci}$ are interestingly $c$-independent; indeed, the notation ``$ci$" stands for this fact. Note that, the gauge solutions appears coupled to the scalar part $H^2 D_2D_1 Z\cdot K$.

\subsection{The General Structure; $c\neq2/5$}
Now, let us study the case $c\neq 2/5$, for which there exists no solution $\widetilde{K}^{g}$ inside $E$. To have a solution, it is necessary to add an extra term $\widetilde{K}'^{g}$ to $\widetilde{K}^{g}$, so that
\begin{eqnarray}
(Q_1 + 6) \widetilde{K}'^{g} = \frac{2-5c}{3(1-c)}\Lambda^\circ,
\end{eqnarray}
where $\Lambda^\circ$ fulfills Eq. (\ref{lambda0}). Therefore, it is obvious that
\begin{eqnarray}
(Q_1 + 6)^2 \widetilde{K}'^{g} = 0.
\end{eqnarray}
$ \widetilde{K}'^{g}$ then can be written as
\begin{eqnarray}
\widetilde{K}'^{g} = \frac{2-5c}{3(1-c)} (Q_1 + 6)^{-1} \Lambda^\circ.
\end{eqnarray}
Accordingly, the general solution for $c\neq2/5$ is
\begin{eqnarray} \label{cd;c}
{\cal K}^{cd;c\neq\frac{2}{5}}=\theta\phi_1 +{\cal S}\bar{Z}K\hspace{3.5cm} \nn\\
+ \frac{2}{27} H^2 D_2D_1 Z\cdot K + D_2( \widetilde{K}^{2g} + \Lambda'^g),\;
\end{eqnarray}
with
\begin{eqnarray}
\Lambda'^g &=& \Lambda^g + \frac{2-5c}{3(1-c)} (Q_1 + 6)^{-1}\nn\\
&&\times \Big( TZ\cdot \bar\partial K -\frac{1}{3} H^2 D_1Z\cdot K + 3 H^2x\cdot ZK \Big).\hspace{0.7cm}
\end{eqnarray}

It is worth mentioning that the term $[(2-5c)/(1-c)](Q_1+6)^{-1}H^2x\cdot ZK$ is responsible for the appearance of logarithmic divergences in the theory. To see the point, let us take a close look at this term. A general solution of $K$ is a linear combination of two scalar fields \cite{Gazeau5920}
\begin{eqnarray}\label{gsK}
K=\bar{Z^\prime}\phi_2+D_1\phi_3,
\end{eqnarray}
where $Z^\prime$ is another constant five-vectors. By inserting (\ref{gsK}) into (\ref{s2}) and using the condition $\bar{\partial}\cdot K=0$ and the following identities
\begin{eqnarray}\label{id11}
Q_1D_1\phi_3=D_1Q_0\phi_3,
\end{eqnarray}
\begin{eqnarray}\label{id22}
Q_1\bar{Z^\prime}\phi_2=(\bar{Z^\prime}(Q_0-2)-2H^2D_1x\cdot Z^\prime )\phi_2,
\end{eqnarray}
we have
\begin{eqnarray}\label{ks1}
Q_0\phi_2=0,
\end{eqnarray}
\begin{eqnarray}\label{ks2}
\phi_3=-(1/2)(Z^\prime\cdot\bar{\partial}+2H^2x\cdot Z^\prime )\phi_2.
\end{eqnarray}
Consequently, the general solution (\ref{gsK}) can be expressed as follows
\begin{eqnarray}\label{kphi}
K=[\bar{Z^\prime}-(1/2)D_1(Z^\prime\cdot\bar{\partial}+2H^2x\cdot Z^\prime)]\phi_2,
\end{eqnarray}
in which, according to (\ref{ks1}), $\phi_2$ corresponds to a MCS field. Here and subsequently, for simplicity of notation, we write $\phi$ instead of $\phi_2$.

The solution to (\ref{ks1}) can be written in terms of the so-called dS massless waves \cite{BGM,BM}
\begin{eqnarray}\label{dsw}
\phi (x)=(Hx\cdot \xi )^\sigma, \;\;\;\;\; \sigma =0,-3
\end{eqnarray}
where this 5-vector $\xi$ lies on the positive null cone ${\cal C}^{+} = \{ \xi \in \mathbb{R}^5;\;\;\xi^2=0,\; {\xi}^0>0 \}$. The vector field $K$, (\ref{kphi}), then takes the following form
\begin{eqnarray}\label{dswk}
K=-\frac{\sigma}{2}\Big[\bar{Z^\prime}+\Big((\sigma -1)\frac{Z^\prime\cdot\xi}{(Hx\cdot\xi)^2}\hspace{2cm}\nn\\
+(\sigma +2)\frac{x\cdot Z^\prime}{x\cdot\xi}\Big)\bar{\xi}\Big]\phi.
\end{eqnarray}

We now turn to the term $(Q_1+6)^{-1}H^2x\cdot ZK$ in (\ref{cd;c}). Utilizing (\ref{dswk}) and (\ref{inv3}) and imposing $Z^\prime\cdot\xi=0$ (for simplicity), one can show that
\begin{eqnarray}\label{invert}
(Q_1+6)^{-1}H^2x\cdot ZK=-(1/2\sigma)H^2x\cdot ZK \hspace{1cm}\nn\\
+(1/2)(Q_1+6)^{-1}f,\hspace{1cm}
\end{eqnarray}
where
$$f\equiv H^2\Big((\sigma +3)x\cdot Z^\prime \bar{Z} +(\sigma +2)H^{-2}\frac{Z\cdot Z^\prime}{x\cdot\xi}\bar{\xi}\Big)\phi.$$
Actually, the first term in the right-hand side of (\ref{invert}) bears a singularity for $\sigma =0$. This implies that the massless spin-2 particles propagate in the interior of the light cone. Of course, setting $c=2/5$, this singularity can be eliminated.

Here, it is interesting to clarify the relationship between the general case, ${\cal K}^{cd;c\neq\frac{2}{5}}$, and the simplest one, ${\cal K}^{cd;c=\frac{2}{5}}$. In the case $c\neq2/5$, one can easily show that
\begin{eqnarray}\label{conphy''}
\partial_2\cdot {\cal K}^{cd;c\neq\frac{2}{5}} &=&\frac{1}{1-c} \Big( TZ\cdot \bar\partial K -\frac{1}{3}H^2 D_1 Z\cdot K \nn\\
& & \hspace{3cm} + 3H^2x\cdot ZK \Big) \nn\\
&=& \frac{1-\frac{2}{5}}{1-c} \Big( \partial_2\cdot {\cal K}^{cd;c=\frac{2}{5}}\Big).
\end{eqnarray}
Combining (\ref{conphy''}) with (\ref{cd;c}), we have
\begin{eqnarray}\label{the}
{\cal K}^{cd;c\neq\frac{2}{5}}&=&  {\cal K}^{cd;c=\frac{2}{5}} \nn\\
&& + \frac{c-\frac{2}{5}}{c-1} D_2 (Q_1+6)^{-1} \partial_2\cdot {\cal K}^{cd;c=\frac{2}{5}}.
\end{eqnarray}
Considering (\ref{ci;''}) and the (\ref{the}), once again, reveals that the divergencelessness solutions are $c$-independent, and interestingly, no logarithmic divergence appears in these solutions. This means the propagation of these modes is confined to the light cone.

We end this section by noting that the above procedure to obtain the general solution to the dS massless spin-2 field equation (\ref{FE}) first developed in \cite{Gazeau2533} for the fields with arbitrary integral spin in Anti-dS spacetime. In our study, however, quite contrary to its Anti-dS counterpart, the invariance of the solutions under the action of the isometry group cannot be preserved in the usual manner utilizing the ordinary positive frequency solutions (with respect to the conformal time). The difference is indeed lied behind the behavior of the structure function, the dS MCS field $\phi$ (see Eq. (\ref{kphi})). This is the subject of our discussion in the next section.

\section{The KGB Structure}
Thus far, the general solution to the field equation (\ref{FE}) has been given. We have shown that the simplest structure would appear in the case $c=2/5$, for which no logarithmic divergent term appears. Of course, the divergencelessness solutions are $c$-independent, and for any choice of the gauge-fixing parameter $c$, they are free of logarithmic divergences.

From now on, for the sake of simplicity, we work with the simplest structure, for which, the formula (\ref{ci;''}) conveys that there exists a corresponding chain in the space of solutions of (\ref{FE}), \emph{i.e.},\footnote{We here simplify our notations by considering ${\cal K}^{cd;c=\frac{2}{5}} \equiv {\cal K}^{cd}$ and $V^{cd;c=\frac{2}{5}} \equiv V^{cd}$.}
\begin{eqnarray}
V^g&=&span\{{\cal K}^g\}\subset V^{ci},\nn\\
V^{ci}&=&span\{{\cal K}^{ci}\}\subset V^{cd},\nn\\
V^{cd}&=&span\{{\cal K}^{cd}\}.
\end{eqnarray}
The last term in (\ref{ci;''}) belongs to the quotient space $V^{cd}/V^{ci}$. We will see in the following that the physical modes lie among the quotient space $V^{ci}/V^g$ (the central part).

In this section, we study the behavior of each of these three spaces of the solutions under the action of the dS group. On this basis, we present a Gupta-Bleuler type structure, which remarkably provides a causal, dS and gauge covariant quantization of the massless spin-2 field.

\subsection{The Emergence of a Krein Space}
We begin our study with the space of the gauge solutions. Under the action of the dS group, the elements of $V^g$ transform as follows
\begin{eqnarray} \label{K'}
L^{(2)}_{\alpha\beta} {\cal{K}}^g = L^{(2)}_{\alpha\beta} D_2\Lambda^g = D_2 L^{(1)}_{\alpha\beta} \Lambda^g.
\end{eqnarray}
Using (\ref{auxi1}), one can easily show that (\ref{K'}) verifies the divergencelessness condition as it is expected for any gauge solutions. The point is that, $L^{(1)}_{\alpha\beta}$ commutes with $Q_1$. On the other hand, putting (\ref{K'}) into the field equation (\ref{FE}), one can also show that it leads to the same equation as (\ref{GS}). These facts show that the subspace of gauge solutions, $ V^g$, is invariant.

The preceding discussion is actually different for the solutions associated with the the divergencelessness part, ${\cal K}^{{ci}} \in V^{ci}$. They satisfy the following equation
\begin{eqnarray}\label{cepaeq}
(Q_2+6){\cal K}^{{ci}}=0.
\end{eqnarray}
Considering ${\cal K}^{{ci}}$ in (\ref{ci;''}) combined with (\ref{kphi}), one can describe these solutions in terms of the polarization tensor ${\cal D}^{{ci}} $ acting on the MCS field $\phi$,
\begin{eqnarray}\label{phypart}
{\cal K}^{ci}= {\cal D}^{{ci}}\phi,
\end{eqnarray}
where
\begin{eqnarray}\label{polphy}
{\cal D}^{{ci}}=&&\Big(-\frac{2}{3}\theta Z\cdot +{\cal S}\bar{Z}+\frac{H^2}{3}D_2[x\cdot Z + \frac{1}{9}D_1Z\cdot ]\Big) \nn\\
&&\times \Big(\bar{Z^\prime}-\frac{1}{2}D_1[Z^\prime\cdot\bar{\partial}+2H^2x\cdot Z^\prime]\Big).\hspace{1.3cm}
\end{eqnarray}
Under the action of the dS group, therefore, they simply transform as follows
\begin{eqnarray} \label{action1}
L^{(2)}_{\alpha\beta} {\cal K}^{ci}_{\gamma\delta} = (L^{(2)}_{\alpha\beta} {\cal D}^{ci}_{\gamma\delta})\phi + {\cal D}^{ci}_{\gamma\delta} (M_{\alpha\beta}\phi).
\end{eqnarray}
It is trivial that the first term on the right-hand side remains invariant. Indeed, since $L^{(2)}_{\alpha\beta}$ commutes with $Q_2$, $(L^{(2)}{\cal D}^{ci})\phi$ is a solution to Eq. (\ref{cepaeq}) as well. Moreover, this term fulfills the divergencelessness condition. One can easily check this fact through the following identity
\begin{eqnarray}\label{intertw}
\partial_2\cdot \Big(L^{(2)} {\cal{D}}^{ci}\Big)\phi = L^{(1)} \Big( \partial_2\cdot {\cal{D}}^{ci}\Big)\phi.
\end{eqnarray}

The second term, however, needs to be evaluated more precisely. In this regard, it is convenient to utilize the bounded global intrinsic coordinates known as conformal coordinates ($X^\mu, \; \mu=0,1,2,3$),
\begin{eqnarray} \label{coordinate}
X=(X^0=H^{-1}\tan\rho, \; (H\cos\rho)^{-1} \Omega)\equiv (\rho, \Omega),
\end{eqnarray}
with $\Omega\in S^3$ and $-\pi/2<\rho<\pi/2$. This system is suitable to describe the compactified dS $\simeq$ Lie sphere $S^3\times S^1$.

Respecting the conformal coordinates and the field equation (\ref{FE}), the following dS-invariant bilinear form (or inner product) can be defined
\begin{eqnarray} \label{B1}
\langle{{\cal{K}}_1},{{\cal{K}}_2}\rangle= \frac{i}{H^2}\int_{S^3,\rho=0} [({{\cal{K}}_1})^*\cdot\cdot\partial_\rho{{\cal{K}}_2} \hspace{2cm}\nn\\
- \frac{4}{5}((\partial_\rho x)\cdot{({{\cal{K}}_1})}^*)\cdot(\partial\cdot{{\cal{K}}_2}) - (1^* \leftrightharpoons 2)]d\Omega,
\end{eqnarray}
where ${\cal{K}}_1$ and ${\cal{K}}_2$ are two arbitrary modes. The above inner product becomes $c$-independent and Klein-Gordon-like if the field verifies the divergencelessness condition
\begin{eqnarray}\label{B2}
({\cal{K}}_1,{\cal{K}}_2 )= \frac{i}{H^2}\int_{S^3,\rho=0} [({\cal{K}}_1)^*\cdot\cdot\partial_\rho{\cal{K}}_2\hspace{1.5cm}\nn\\
- {\cal{K}}_2\cdot\cdot\partial_\rho({\cal{K}}_1)^*]d\Omega.
\end{eqnarray}

Now, according to (\ref{666}), the intrinsic counterpart of the solution ${\cal K}^{ci}$ would be
\begin{eqnarray}\label{first}
{\cal K}^{ci}_{\mu\nu}(X)=\Delta^{ci}_{\mu\nu}(\rho,\Omega,Llm) \phi_{Llm}(\rho,\Omega)\equiv {\cal K}^{{ci};Llm}_{\mu\nu},
\end{eqnarray}
with $ \Delta^{ci}_{\mu\nu} = x^\alpha_\mu x^\beta_\nu {\cal{D}}^{ci}_{\alpha\beta}$. The ``strictly positive" solutions to the structure function, the dS MCS field, are given by
\begin{eqnarray} \label{phi+}
\phi_{Llm} = \chi_L(\rho) {\cal{Y}}_{Llm}(\Omega),
\end{eqnarray}
with $L=1,2,..., \;\; 0\leq l\leq L,\;\; 0\leq |m|\leq l$. The ${\cal{Y}}_{Llm}$ are the spherical harmonics on $S^3$. The $ \chi_L(\rho) $ are obtained by the massless limit of the usual Bunch-Davies modes \cite{Birrell}
\begin{eqnarray} \label{chi}
\chi_L(\rho) = A_L(Le^{-i(L+2)\rho}+(L+2)e^{-iL\rho}),
\end{eqnarray}
where $A_L= \frac{H}{2}[2L(L+1)(L+2)]^{-1/2}$.

The modes (\ref{first}) form an orthonormal system with respect to the inner product (\ref{B2}). Take a close look at the above formula, however, reveals that the normalization constant $A_L$ breaks down at $L=0$; this is known as the ``zero-mode" problem \cite{AF}. This difficulty arises due to the fact that the set constructed over the strictly positive modes is not invariant under the action of the dS group \cite{Gazeau1415,de Bievre6230}. If one insists on the full dS invariance, it would be required to deal with the $L=0$ solutions. There are two of them, $\phi^{(1)}_{0,0,0}$ and $\phi^{(2)}_{0,0,0}$, where
\begin{eqnarray} \label{phigz}
&&\phi^{(1)}_{0,0,0}= \mbox{\emph{constant}}= \frac{H}{2\pi},\nn\\
&&\phi^{(2)}_{0,0,0} = -i\frac{H}{2\pi} (\rho + \frac{1}{2}\sin 2\rho).
\end{eqnarray}
However, both are null norm modes, with respect to the natural Klein-Gordon inner product associated with the MCS field (we refer to this inner product as $(\;,\;)_{{MCS}}$). To fix this degeneracy, one should define \cite{Gazeau1415,de Bievre6230}
\begin{eqnarray} \label{000}
\phi_{0,0,0}= \phi^{(1)}_{0,0,0} + \phi^{(2)}_{0,0,0} / {2}.
\end{eqnarray}
It is the ``true zero-mode" of Allen \cite{AF}. The constants of normalization are chosen to have $( \phi_{0,0,0},\phi_{0,0,0})_{{MCS}}=1$.

Considering this mode, we have a complete set of strictly positive norm modes for $L\geq0$, but the space constructed over these modes is not dS invariant; under the dS group actions, the zero-mode produces negative modes ($\phi_{Llm}^\ast$) as well as positive modes ($\phi_{Llm}$). Indeed, if we consider the following categories of the space of solutions, $\phi_{Llm,L>0}\in {\cal{V}}^+$, $\phi_{Llm,L>0}^\ast\in {\cal{V}}^-$, $\phi^{(1)}_{0,0,0}\in {\cal{N}}$ and $\phi^{(2)}_{0,0,0}\in {\cal{M}}$, under the action of the dS group, we have \cite{Gazeau1415,de Bievre6230}
\begin{eqnarray} \label{transform}
&&{U}_g: {\cal{N}} \rightarrow {\cal{N}},\nn\\
&&{U}_g: {\cal{M}} \rightarrow {\cal{M}}\oplus {\cal{V}}^+\oplus {\cal{V}}^-\oplus {\cal{N}},\nn\\
&&{U}_g: {\cal{V}}^+ \rightarrow {\cal{V}}^+\oplus {\cal{N}},\nn\\
&&{U}_g: {\cal{V}}^- \rightarrow {\cal{V}}^-\oplus {\cal{N}}.
\end{eqnarray}
For any $g$ in the dS group, $U_g$ stands for the dS natural representation on the space of solutions. Therefore, it seems that the smallest, complete, non-degenerate, and invariant inner product space for the MCS field would be ${\cal{V}}^+ \oplus {\cal{V}}^- \oplus {\cal{N}} \oplus {\cal{M}}$.

It is obvious that the same argument appears for the general solutions. The invariance of the divergencelessness space and also the total space of solutions, therefore, inevitably necessitates extending $V^{cd}$ to a Krein space, which includes all the negative frequency solutions to the field equation (\ref{FE}),
\begin{eqnarray}\label{krein space}
V^{cd} = {\cal{H}} \oplus {\cal{H}}^\ast,
\end{eqnarray}
where
\begin{eqnarray}\label{set1}
{\cal{H}}  = \{\sum_{Llm,\; L\geq0} c_{Llm} {\cal{K}}^{{cd};Llm}_{\mu\nu};  \sum_{Llm,\; L\geq0} |c_{Llm}|^2<\infty\}.\;
\end{eqnarray}

Accordingly, we have a chain of invariant subspaces $V^g\subset V^{ci}\subset V^{cd}$ carry an indecomposable group representation structure; a Krein-Gupta-Bleuler triplet. The gauge solutions are orthogonal to the ones belong to $V^{ci}$ including themselves. They constitute the invariant subspace $V^g$ which is not invariantly complemented in $V^{ci}$. There is a similar situation for the divergencelessness solutions. They build up the invariant subspace $V^{ci}$ which is not invariantly complemented in $V^{cd}$. In summary, the indecomposable group representation structure associated with the massless spin-2 field is simply demonstrated as follows
\begin{eqnarray}\label{trirep}
\underbrace{\Pi_{2,0}}_{V^{cd}/V^{ci}}\;\;\longrightarrow\underbrace{\Pi_{2,2}^{+}
\oplus\Pi_{2,2}^{-}}_{V^{ci}/V^g}\longrightarrow\;\;\underbrace{\Pi_{2,0}}_{V^g}
\end{eqnarray}
Note that, from Eq. (\ref{K'}), one can easily conclude that the gauge solutions ${\cal{K}}^g = D_2 \Lambda^g$ are completely characterized by $ \Lambda^g $, which respecting Eq. (\ref{GS}) obeys $(Q_1+6)\Lambda^g = (Q_0+4)\Lambda^g=0 $ (since $x\cdot\Lambda^g= \bar\partial\cdot\Lambda^g=0$). The gauge solutions, therefore, can be associated with $\Pi_{2,0}$. The same argument appears for the solutions belong to $V^{cd}/V^{ci} $. This is indeed the point lying behind the above formula.

\subsection{The Physical Subspace}
With respect to the above statements, the space of the physical modes (denoted by ${\cal{H}}^p$) of the massless spin-2 field would be
\begin{eqnarray}\label{phys states}
{\cal{H}}^p = \{ \sum_{Llm,\; L>0} c_{Llm} {\cal{K}}^{\frac{ci}{g};Llm}_{\mu\nu}; \sum_{Llm,\; L>0} |c_{Llm}|^2<\infty\},\;
\end{eqnarray}
in which ${\cal{K}}^{\frac{ci}{g};Llm}_{\mu\nu} \in V^{ci}/V^g$. Regarding this definition, it is obvious that the physical modes are $c$-independent. They propagate only on the dS light cone. Moreover, ${\cal{H}}^p$ is closed under the action of the dS group. In this regard, it must be underlined that the massless spin-2 mode with $L=0$, does not belong to the space of the physical solutions, because if it was considered, the subspace of the positive norm modes would be transformed into the subspace of the negative norm modes violating unitarity (see (\ref{transform})). Indeed, when we study any physical quantity, only the strictly positive solutions with $L>0$ are taking into account.

\begin{widetext}
\section{The Quantum Field}
In this section, we proceed with the KGB quantization of massless spin-2 field in dS space. Indeed, we define a new representation of the canonical commutation relations which lead to a covariant field. This field, as expected, is a distribution for which the values are operators on the bosonic Fock space constructed upon the total space (see section. IV and \cite{Mintchev} for a review of the theory of Fock spaces on Krein spaces).

Before we go further, let us simplify the previous notation by means of the following definition
\begin{eqnarray}
{\cal J}=\{j\equiv (L,l,m)\in \mathbb{N}\times\mathbb{N}\times\mathbb{Z};\;L\geq 0,\; 0\leq l\leq L,\; -l\leq m\leq l\}.
\end{eqnarray}
Therefore, from now on, we have ${\cal K}^{cd;Llm}\equiv {\cal K}^{cd;j}$.

In the KGB structure, given any $V^{cd}$, we denote by $\underline{V}^{cd}$ the corresponding Fock space, for which, the annihilator of a solution ${\cal{K}}^{cd}$ is defined by
\begin{eqnarray}\label{ann}
\Big(a({\cal{K}}^{cd;j})\Psi\Big)(X_1,...,X_{n-1})=\sqrt{n} \frac{i}{H^2}\int_{\rho=0}\Big[({\cal{K}}^{cd;j})^\ast(\rho,\Omega)\cdot\cdot\partial_\rho{\Psi}\Big((\rho,\Omega),X_1,...,X_{n-1}\Big)\hspace{4cm}\nn\\
- \frac{4}{5}((\partial_\rho x)\cdot({{\cal{K}}}^{cd;j})^\ast(\rho,\Omega))\cdot(\partial\cdot{\Psi})\Big((\rho,\Omega),X_1,...,X_{n-1})\Big) - (1^\ast \leftrightharpoons 2)\Big]d\Omega,
\end{eqnarray}
for any square-integrable $n$-symmetric function $\Psi$. As usual, we define the creator by
\begin{eqnarray}\label{crea}
\Big(a^\dagger({\cal{K}}^{cd;j})\Psi\Big)(X_1,...,X_{n+1})= \frac{1}{\sqrt{n+1}} \sum_{i=1}^{n+1} {{\cal{K}}}^{cd;j}(X_i)\cdot\cdot{\Psi}(X_1,...,\widetilde{X}_i,...,X_{n+1}),
\end{eqnarray}
where $\widetilde{X}_i$ indicates the omission of this term.

We can now define the quantum field $\underline{\cal{K}}^{cd}_{\mu\nu}(X)$ on $\underline{V}^{cd}$ by
\begin{eqnarray}\label{qf}
\underline{\cal{K}}^{cd}_{\mu\nu}(X) = \sum_{j} a_j {\cal{K}}_{\mu\nu}^{cd;j} + \sum_{j} a_j^\dagger ({\cal{K}}_{\mu\nu}^{{cd};j})^\ast - \sum_{j} b_j ({\cal{K}}_{\mu\nu}^{cd;j})^\ast - \sum_{j} b_j^\dagger {\cal{K}}_{\mu\nu}^{cd;j},
\end{eqnarray}
in which $a_j\equiv a({\cal K}^{cd;j})$ and $b_j\equiv a(({\cal K}^{cd;j})^\ast)$ are, respectively, the annihilators of the modes ${\cal{K}}_{\mu\nu}^{cd;j}$ and $({\cal{K}}_{\mu\nu}^{cd;j})^\ast$. For any ${\cal K}_1,{\cal K}_2\in V^{cd}$, these operators obey the following commutation relation
\begin{eqnarray} \label{causality}
[a({\cal{K}}_1),a^\dagger({\cal{K}}_2)]=\langle {\cal{K}}_1,{\cal{K}}_2\rangle,
\end{eqnarray}
and we have
\begin{eqnarray}\label{gives}
\underline{U}_g a({\cal{K}}_1) \underline{U}_g^{\ast} = a(U_g{\cal{K}}_1), \;\;\; \underline{U}_g a^\dagger({\cal{K}}_1) \underline{U}_g^\ast = a^\dagger(U_g{\cal{K}}_1).
\end{eqnarray}
\end{widetext}
Note that, $\underline{U} $ is the extension of the natural representation $U$ of the dS group on $V^{cd}$ to the Fock space $\underline{V}^{cd}$.

At this point, we would like to investigate the causality and covariance of the above quantum field. Actually, for any real test function $f_{\mu\nu}$ in the space of functions $C^\infty$ with compact support in ${M_{H}}$, there exists a unique element $p(f)\in V^{cd}$ for which
\begin{eqnarray}\label{uniqelem}
p_{\mu\nu}(f)=\int_{{M_{H}}}{\cal K}(f){\cal K}_{\mu\nu}(X)d\sigma(X),
\end{eqnarray}
where $d\sigma(X)$ is the dS-invariant measure and the smeared form of the modes  ${\cal K}(f)$ is
\begin{eqnarray}\label{smearout}
{\cal K}(f)=\int_{{M_{H}}}{\cal K}^\ast_{\mu\nu}(X)f^{\mu\nu}(X)d\sigma(X).
\end{eqnarray}
From (\ref{smearout}) and noticing this fact that the space of solutions $V^{cd}$ is non-degenerate and invariant, one can immediately concludes that
\begin{eqnarray}\label{commu}
U_gp(f)=p(U_gf).
\end{eqnarray}

On this basis, the smeared field is
\begin{eqnarray}\label{smfield}
\underline{{\cal K}}^{cd}(f)=a(p(f))+a^\dagger(p(f)),
\end{eqnarray}
and its unsmeared form becomes
\begin{eqnarray}\label{phixfi}
\underline{{\cal K}}^{cd}(X)=a(p(X))+a^\dagger (p(X)).
\end{eqnarray}
We are now in the position to check the covariance of the quantum field. Indeed, we have
\begin{eqnarray}
\underline{U}_g\underline{{\cal K}}^{cd}(X)\underline{U}_g^{-1}&=&a(U_gp(X))+a^\dagger (U_gp(X))\nn\\
&=&a(p(g\cdot X))+a^\dagger (p(g\cdot X))\nn\\
&=&\underline{{\cal K}}^{cd}(g\cdot X),
\end{eqnarray}
where we use (\ref{gives}) and the unsmeared form of (\ref{commu}).

Let us now investigate the causality of the theory. The kernel of the distribution $p$ is the so-called propagator $\tilde{G}_{\mu\nu\mu^\prime\nu^\prime} =G^{adv}_{\mu\nu\mu^\prime\nu^\prime} -G^{ret}_{\mu\nu\mu^\prime\nu^\prime}$, that is to say
$$\langle p_{\mu\nu}(X), p_{\mu^\prime\nu^\prime}(X^\prime)\rangle=-i\tilde{G}_{\mu\nu\mu^\prime\nu^\prime}(X,X^\prime).$$
Using (\ref{commu}), one can easily show that the propagator is invariant under the action of the dS group. Furthermore, one finds directly the commutation relation between the fields as follows
\begin{eqnarray}\label{fcommu}
[\underline{{\cal K}}_{\mu\nu}^{cd}(X), \underline{{\cal K}}_{\mu^\prime\nu^\prime}^{cd}(X^\prime)]&=&2\langle p_{\mu\nu}(X), p_{\mu^\prime\nu^\prime}(X^\prime)\rangle \nn\\
&=&-2i\tilde{G}_{\mu\nu\mu^\prime\nu^\prime}(X,X^\prime).\hspace{.1cm}
\end{eqnarray}
As a consequence, the fields satisfy causal commutation relation because $\tilde{G}$ vanishes when $X$ is spacelike separated from $X^\prime$.

Now, we can accurately identify the KGB Fock vacuum as follows
\begin{equation}
a_j |0\rangle = b_j |0\rangle = 0,\;\; \forall j\in {\cal{J}}.
\end{equation}
It is obviously dS invariant.

At the end, it must be underlined that the KGB vacuum does not depend on Bogolubov transformations which merely modify the set of physical states. This is not however surprising, because in our formalism not only is the vacuum different but so is the field itself. We do insist here, this does not imply that Bogolubov transformations are not valid any more. Indeed, under the Bogolubov transformations the space ${\cal{H}}=\mbox{\emph{span}}({\cal{K}}^{{cd};j}_{\mu\nu})$ would transform to $\widetilde{{\cal{H}}} = \mbox{\emph{span}}(\widetilde{{\cal{K}}}^{{cd};j}_{\mu\nu})$, based upon which we have a new representation for the first two terms on the right hand side of (\ref{qf}). The crucial point, however, is that the total space and correspondingly the field representation remain unchanged,
$$\widetilde{{\cal{H}}} \oplus \widetilde{{\cal{H}}}^\ast = {{\cal{H}}} \oplus {{\cal{H}}}^\ast,\;\;\; \underline{\widetilde{{\cal{K}}}}^{{cd};j}_{\mu\nu}= \underline{{{\cal{K}}}}^{{cd};j}_{\mu\nu}.$$
Again, our quantization scheme is of Gupta-Bleuler type, in the sense that one should distinguish the total space from the subspace of the physical states based upon which, respectively, the observables and the mean values of them are determined. This is the subject of our discussion in the following section.

\section{Physical Content of the Theory}
Thus far, the quantum field has been constructed which is causal and has all the covariance properties of the classical field. The price to pay is, however, the presence of some non-physical states in our construction. In order to have a meaningful interpretation, therefore, it is essential to select the subspace of physical states and also prove that the presence of non-physical states will not result in any negative energies. These issues are discussed in detail in this section.

We first start by identifying the gauge states space. The space of the dS-invariant states of $\underline{V}^{cd}$ is $\underline{V}^{g}$, the space defined by $ a^\dagger_g|0\rangle$ ($a^\dagger_g \equiv a^\dagger ({\cal{K}}^g)$). We denote by $\underline{V}^{ci}$ the corresponding subspace of divergenceless states which is generated from the Fock vacuum by $(a^\dagger_g)^{n_0}(a^\dagger_{j_1})^{n_1} \cdot\cdot\cdot (a^\dagger_{j_l})^{n_l} |0\rangle$. Here, we also designate by $\underline{\underline{V}}^g$ the subspace of $\underline{V}^{ci}$ orthogonal to $\underline{V}^{ci}$, $\underline{\underline{V}}^g = \underline{V}^{ci} \cap (\underline{V}^{ci})^\bot$, so that, $\underline{V}^{g}\subset\underline{\underline{V}}^g$. $\underline{\underline{V}}^g$ is indeed the set of unobservable gauge states and defined as follows
\begin{eqnarray}\label{Vg2}
\textbf{{\cal{K}}} \in \underline{\underline{V}}^g \;\;\; \mbox{if} \;\;\; \textbf{{\cal{K}}} \in \underline{V}^{ci} \;\;\; \mbox{and} \;\;\; (\textbf{{\cal{K}}},\textbf{{\cal{K}}}') = 0 \;\;\; \forall \textbf{{\cal{K}}}' \in \underline{V}^{ci}. \;\;
\end{eqnarray}
For any state $\textbf{{\cal{K}}} \in \underline{V}^{ci}$, the state $a^\dagger_g \textbf{{\cal{K}}} \in \underline{\underline{V}}^g$; these two states are equal up to an element of $\underline{\underline{V}}^g$. So, including the Fock space $\underline{V}^{cd}$ built on $V^{cd}$, the second-quantized Gupta-Bleuler triplet is obtained
$$\underline{\underline{V}}^g \subset \underline{V}^{ci} \subset \underline{V}^{cd},$$
that is apparently invariant under the de Sitter group action.

With respect to the definition of the dS-invariant space $\underline{V}^g$ presented above, it is of infinite dimension subspace of $\underline{\underline{V}}^g$. As a result, it seems that the Fock vacuum is not the only dS-invariant state. However, it is crucial to refer the reader to the definition of physical equivalence, according to which, all these states are equal to an element of that dimensional space. This is indeed called Quasi-uniqueness of the KGB Fock vacuum and understood that the vacuum transforms into a physically equivalent state under gauge transformation (simply, the vacuum is gauge invariant).

We maintain that the physical states belong to the central space $\underline{V}^{ci} / \underline{\underline{V}}^g$, where a gauge transformation maps an element into an equivalent element of the central space. Nonetheless, in order to determine the subspace of physical states, as pointed out in the previous sections, we require to impose an extra condition; due to the structure function of the theory, the MCS field, some states in the central part have negative norm. One has to exclude these states (see (\ref{phys states}) and the associated explanations) to obtain the true physical states space, more precisely, the Hilbert space carrying the physical representation of the massless spin-2 field equipped with positive invariant inner product. We denote this space by $\underline{\cal H}^p$. Two physical states $\cal {P}$ and ${\cal {P}}'$ are called equivalent if ${\cal{P}} - {\cal{P}}' \in \underline{\underline{V}}^g$. Physical states are particularly important since in a Gupta-Bleuler formalism mean values of observables, which will be defined below, are determined by them.

On this basis, an observable $O$ (\emph{e.g.} the stress tensor $T_{\mu\nu}$) is a symmetric operator on $\underline{V}^{cd}$ such that for two equivalent physical states $\cal {P}$ and ${\cal {P}}'$ defined above, we must have
$$\langle {\cal{P}} |O| {\cal{P}} \rangle = \langle {\cal{P}}' |O| {\cal{P}}' \rangle.$$
This means that expectation values of observables are gauge independent. In this regard, it is not difficult to show that the field $\underline{\cal{K}}^{cd}$ itself does not justify the definition of observables. This fact directly results in gauge dependency of two-point functions which are written in terms of the field, such as Wightman or Hadamard functions,
$$\langle 0|\underline{{\cal{K}}}(X) \underline{{\cal{K}}}(X') |0 \rangle, \;\; \langle 0|\underline{{\cal{K}}}(X)\underline{ {\cal{K}}}(X') +  \underline{{\cal{K}}}(X') \underline{{\cal{K}}}(X)|0 \rangle,$$
Therefore, in this quantization scheme, it is not expected that the symmetric two-point function, Hadamard function, has significant physical interpretation and a simple calculation reveals that it vanishes. Of course, it is a direct consequence of demanding the full dS covariance of the theory, which unavoidably necessitates negative norm states (the KGB structure).\footnote{A straightforward calculation reveals that the only full dS-covariant and causal two-point function which naturally appears is the commutator, but it is not of positive type \cite{II,I} (see also \cite{Bamba}).} In this construction, therefore, two-point functions and the vacuum are not linked as the standard QFT; as opposed to the usual QFT for which choosing a vacuum is equivalent to choosing a physical states space and a two-point function, in this context, the KGB vacuum is unique and cannot identify the space of physical states. This space, however, is still linked to the two-point functions \cite{Gazeau1415}. This means that a two-point function with Hadamard property is available but with another meaning (for electromagnetic field on globally hyperbolic spacetimes, see \cite{Strohmaier,Wrochna}).

Now, we show that although the stress tensor is defined on the total space which includes negative norm states, no negative energy would be yielded; in general, we have
\begin{eqnarray}\label{qqq}
\langle 0|T_{\mu\nu}|0 \rangle &=& \sum_{j\in{\cal{J}}} T_{\mu\nu} [{\cal{K}}_{\mu\nu}^{cd;j},({\cal{K}}_{\mu\nu}^{cd;j})^\ast]\nn\\
&-& \sum_{j\in{\cal{J}}}T_{\mu\nu} [({\cal{K}}_{\mu\nu}^{cd;j})^\ast, {\cal{K}}_{\mu\nu}^{cd;j}] = 0,
\end{eqnarray}
where $ T_{\mu\nu} [{\cal{K}},{\cal{K}}]$ denotes the bilinear expression of the stress tensor $T_{\mu\nu}$. Note that, the cancellation in (\ref{qqq}) is due to the unusual second term on the right hand side which comes from the terms of the field containing $b_j$ and $b_j^\dagger$.

Similarly we can compute the mean values of the stress tensor on physical states, $|\overrightarrow{{\cal{P}}} \rangle = \frac{1}{\sqrt{n_1! ... n_l!}} (a^\dagger_{j_{({\cal{P}}_1)}})^{n_1} ... (a^\dagger_{j_{({\cal{P}}_l)}})^{n_l} |0 \rangle$,
\begin{eqnarray}\label{EMT1}
\langle \overrightarrow{{\cal{P}}}| T_{\mu\nu} |\overrightarrow{{\cal{P}}} \rangle = 2 Re \sum_{{j=1}}^{{l}} n_j T_{\mu\nu} [{\cal{K}}_{\mu\nu}^{cd;j},({\cal{K}}_{\mu\nu}^{cd;j})^\ast].
\end{eqnarray}
As a consequence we have
\begin{eqnarray}\label{EMT1'}
\langle \overrightarrow{{\cal{P}}}|T_{00}|\overrightarrow{{\cal{P}}} \rangle \geq0.
\end{eqnarray}

The KGB formalism indeed provides an automatic and covariant renormalization of the stress tensor, which remarkably, fulfills the so-called Wald axioms;
\begin{itemize}
\item First, the field is causal and covariant, therefore, the causality and the covariance of the stress tensor are guaranteed.

\item Second, considering the physical states, the formalism gives the formal results.

\item Third, computing the mean values of the stress tensor in the physical states, the procedure is equivalent to reordering. The crucial point here is that
$$[a_j,a_j^\dagger] = -[b_j,b_j^\dagger],$$
which implies
$$ a_ja_j^\dagger + a_j^\dagger a_j +  b_jb_j^\dagger + b_j^\dagger b_j = 2a_j^\dagger a_j + 2b_j^\dagger b_j.$$
\end{itemize}

With respect to the above statements, one can easily see that, considering the KGB quantum field, no trace anomaly appears in the computation of the energy-momentum tensor; the expected value of all components of the stress tensor vanish in the KGB vacuum. Of course this is not very surprising, because the KGB quantization method preserves covariance and conformal covariance of the theory in a rather strong sense, and therefore, the theory does not exhibit any trace anomaly which, after all, can appear only by breaking the conformal invariance.

\section{Summery and Discussion}
In this paper, we have dealt with a subject that has been controversial for a rather long period of time: the quantization of the massless spin-2 (graviton) field in dS space. Together with the quantization of the MCS field in dS space, they form a doublet of cases where quantization can lead to surprises.

The case of the MCS was analyzed by Allen and Folacci \cite{AF} and their results seemed definitive; dS invariance was broken and infrared divergences were present due to the ever-increasing number of modes exiting the horizon. Nevertheless, thanks to a new representation of the canonical commutation relations based on the KGB method, quantization of the MCS field which satisfies full covariance as well as causality has been proposed in \cite{de Bievre6230,Gazeau1415}. It must be emphasized that there is no contradiction with Allen's point of view since in this formalism vacuum and the field itself are different.

The case of the graviton and the associated dS symmetry breaking, however, are more complicated among other things because of the local invariance present, in contradistinction to the scalar case. Debate over this issue has gone on for decades with the particle physics community (see, for instance, \cite{SMiao104004,RWoodard1430020,Miao0,Miao00}) maintaining that gravitons inherit the dS breaking long recognized for the MCS field and the mathematical physics community (see, for instance, \cite{higuchi,Marolf&Morrison,Faizal&Higuchi,higuchi1,Bernar,higuchi2}) maintaining that there is no physical breaking of dS invariance. In this section, we briefly discuss the place of our approach amongst them.

Recently, based on a rigorous group theoretical approach and in consistency with the particle physics community viewpoint, we have shown that there exists no natural dS-invariant vacuum state (the Bunch-Davies state) for the graviton field in dS space \cite{Bamba} and correspondingly the associated infrared divergences cannot be gauged away \cite{I,II}. Indeed, it seems that within the framework of usual QFT, one has to consider a restrictive version of covariance with respect to some maximal subgroup of the dS group only ($SO(4)$, $SO(1,3)$ or $E(3)$). From the perspective of the mathematical physics community to which we belong, however, dS space has a privileged status as the unique, maximally symmetric solution to the Einstein equation with positive cosmological constant. It provides the opportunity of controlling the transition to the flat space by the procedure known as contraction procedure \cite{Garidi124028}. Accordingly, dS space should at least be respected as an excellent laboratory. On the other hand, being the most serious candidate for a complete quantum theory of gravity, string theory should admit dS vacua. Indeed, there are several reasons, such as a full understanding of holography for dS gravity \cite{Hooft} and clarifying the microscopic origin of dS entropy \cite{Gibbons}, that make it desirable to embed dS space in string theory. From this point of view, it seems that, a crucial step to take would be constructing a fully covariant QFT in dS space.

Motivated by all the above reasons and following our previous work \cite{Bamba}, in this paper, we have constructed a causal and dS-covariant (more exactly, $SO_0(1,4)$-covariant) free massless spin-2 quantum field (graviton field) on dS spacetime admitting a dS-invariant vacuum in an indefinite inner product space. Quite similar to the reasoning given in \cite{de Bievre6230,Gazeau1415} for the MCS field, the causality and the covariance of the theory are assured thanks to a suitable adaptation (Krein spaces) of the Wightman-G\"{a}rding axiomatic for massless fields (the Gupta-Bleuler structure). Our KGB quantization scheme is, therefore, free of any infrared divergence. Again, it is indeed because of our choice of the Krein vacuum not the gauge-fixing procedure. Pursuing our quantization scheme, we have also specified the space of physical states. The theory, despite the appearance of the non-physical negative norm states in the quantization procedure, gives the correct sign for the energy on the physical states (note that, the so-called Wald axioms are already well preserved).

Here, it must be underlined that, when interaction is present, with respect to the procedure given in \cite{Garidi2005}, determining the space of physical states is the critical step in defining the unitary condition of the theory. Applying the unitary condition, it is proved that this quantization scheme in Minkowski space when interaction is taken into account truly yields the common results; the so-called radiative corrections are indeed the same as usual QFT (see the mathematical details in \cite{Garidi2005}). It also allows us to obtain the exact usual result for the black hole radiation, even regarding that the free field vacuum expectation value of the energy-momentum tensor is zero \cite{blackhole} (in this regard, see also \cite{braneII,braneI,Casimir}). On the other hand, following Wald, there exists a case where this quantization method seems in a very natural way. According to the statement given by Wald (\cite{WaldBook} p.66): ``\emph{For a spacetime which is asymptotically stationary in both the past and the future we have two natural choices of vacua, and the $S$ matrix should be a unitary operator between both structures}", as he mentioned this is not possible if the two vacua are not equivalent. Remembering the Krein vacuum is unique \cite{Garidi2005}, it seems that the Krein quantization method provides a stage where all these objects can be respected together.

Now, the natural question which arises is that can one formulate perturbative field theory for quantum gravity in dS space through the KGB method? We think it is too early to answer this, and much more work is still necessary, specially in constructing interacting field theory on dS space. As already pointed out, the Hadamard property still needs to be restored somehow (for electromagnetic field on globally hyperbolic spacetimes, see \cite{Strohmaier,Wrochna}). The correctness of this approach will ultimately be decided by experiment and observation.

\section*{Acknowledgements}
This work was partially supported by the JSPS KAKENHI Grant Number JP 25800136 and the research-funds presented by Fukushima University (K.B.).


\begin{thebibliography}{a}

\bibitem{Nojiri:2010wj} S. Nojiri and S.D. Odintsov, Phys. Rept. 505, 59 (2011).

\bibitem{Nojiri:2006ri} S. Nojiri and S.D. Odintsov, Int. J. Geom. Meth. Mod. Phys. 4, 115 (2007).

\bibitem{Book-Capozziello-Faraoni} V. Faraoni and S. Capozziello, \emph{Fundamental Theories of Physics}, (Springer, New York, 2010), Vol. 170.

\bibitem{Capozziello:2011et} S. Capozziello and M. De Laurentiz, Phys. Rept. 509, 167 (2011).

\bibitem{delaCruzDombriz:2012xy} A. de la Cruz-Dombriz and D. S\'{a}ez-G\'{o}mez, Entropy 14, 1717 (2012).

\bibitem{Bamba:2012cp} K. Bamba, S. Capozziello, S. Nojiri, and S.D. Odintsov, Astrophys. Space Sci. 342, 155 (2012).

\bibitem{Joyce:2014kja} A. Joyce, B. Jain, J. Khoury, and M. Trodden, Phys. Rept. 568, 1 (2015).

\bibitem{Koyama:2015vza} K. Koyama, Rept. Prog. Phys. 79, 046902 (2016).

\bibitem{Bamba:2015uma} K. Bamba and S.D. Odintsov, Symmetry 7, 220 (2015).

\bibitem{Nojiri:2017ncd} S. Nojiri, S.D. Odintsov, and V.K. Oikonomou, Phys. Rep. 692, 1-104 (2017).

\bibitem{Dixmier} J. Dixmier, Bull. Soc. Math. France, 89, 9 (1961).

\bibitem{Takahashi289} B. Takahashi, Bull. Soc. Math. France 91, 289 (1963).

\bibitem{Flato415} M. Flato, C. Fronsdal and J.P. Gazeau, Phys. Rev. D 33, 415 (1986).

\bibitem{angelo} E. Angelopoulos, M. Flato, C. Fronsdal, and D. Sternheimer, Phys. Rev. D 23, 1278 (1981).

\bibitem{angeloflato} E. Angelopoulos and M. Flato, Lett. Math. Phys. 2, 405 (1978).

\bibitem{levy} M. Levy-Nahas, J. Math. Phys. 8, 1211 (1967).

\bibitem{barut} A.O. Barut , A. B\"ohm, J. Math. Phys. 11, 2938 (1970).

\bibitem{Gazeau507} J.P. Gazeau, Lett. Math. Phys. 8, 507 (1984).

\bibitem{Gazeau2533} J.P. Gazeau, M. Hans, J. Math. Phys. 29, 2533 (1988).

\bibitem{Gazeau329} J.P. Gazeau, M. Hans, and R. Murenzi, Class. Quant. Grav. 6, 329 (1989).

\bibitem{bacry} H. Bacry, J.M. Levy-Leblond, J. Math. Phys. 9, 1605 (1968).

\bibitem{What} T. Garidi, \emph{What is mass in desitterian physics?}. arXiv preprint hep-th/0309104 (2003).

\bibitem{Garidi3838} T. Garidi, J.P. Gazeau and M.V. Takook, J. Math. Phys. 44, 3838 (2003).

\bibitem{II} H. Pejhan and S. Rahbardehghan, Phys. Rev. D 94, 104030 (2016).

\bibitem{I} H. Pejhan and S. Rahbardehghan, Phys. Rev. D 93, 044016 (2016).

\bibitem{BinegarGUPTA} B. Binegar, C. Fronsdal and W. Heidenreich, J. Math. Phys. 24, 2828 (1983).

\bibitem{Gazeau1847} J.P. Gazeau, J. Math. Phys. (N.Y.) 26, 1847 (1985).

\bibitem{Garidi032501} T. Garidi, J.P. Gazeau, S. Rouhani and M.V. Takook, J. Math. Phys. 49, 032501 (2008).

\bibitem{Gazeau5920} J.P. Gazeau and M.V. Takook, J. Math. Phys. 41, 5920 (2000).

\bibitem{BGM} J. Bros, J. P. Gazeau, and U. Moschella, Phys. Rev. Lett. 73, 1746 (1994).

\bibitem{BM} J. Bros and U. Moschella, Rev. Math. Phys. 08, 327 (1996).

\bibitem{Birrell} N.D. Birrell and P.C.W. Davies, \emph{Quantum fields in curved space}, Cambridge University Press, Cambridge, England, (1984).

\bibitem{AF} B. Allen and A. Folacci, Phys. Rev. D 35, 3771 (1987).

\bibitem{de Bievre6230} S. De Bi\`{e}vre and J. Renaud, Phys. Rev. D 57, 6230 (1998).

\bibitem{Gazeau1415} J.P. Gazeau, J. Renaud, and M.V. Takook, Class. Quant. Grav. 17, 1415 (2000).

\bibitem{Mintchev} M. Mintchev, J. Phys. A: Math. Gen. 13, 1841 (1980).

\bibitem{Bamba} K. Bamba, S. Rahbardehghan and H. Pejhan, Phys. Rev. D 96, 106009 (2017).

\bibitem{Strohmaier} F. Finster, A. Strohmaier, Ann. Henri Poincare 16, 1837 (2015).

\bibitem{Wrochna} M. Wrochna, Ann. Henri Poincaré 13, (2012) 8.

\bibitem{Miao0} S.P. Miao, N.C. Tsamis and R.P. Woodard, J. Math. Physics 52, 122301 (2011).

\bibitem{Miao00} S.P. Miao, N.C. Tsamis and R.P. Woodard, Class. Quantum Grav. 28, 245013 (2011).

\bibitem{SMiao104004} S.P. Miao, P.J. Mora, N.C. Tsamis, and R.P. Woodard, Phys. Rev. D 89, 104004 (2014).

\bibitem{RWoodard1430020} R.P. Woodard, Int. J. Mod. Phys. D 23, 1430020 (2014).

\bibitem{higuchi} A. Higuchi, Class. Quant. Grav. 8, 2005 (1991).

\bibitem{Marolf&Morrison} D. Marolf and I.A. Morrison, Class. Quantum Grav. 26, 235003 (2009).

\bibitem{Faizal&Higuchi} M. Faizal and A. Higuchi, Phys. Rev. D 85, 124021 (2012).

\bibitem{higuchi2} M.B. Fr¨ob, A. Higuchi, and W.C.C. Lima, Phys. Rev. D 93, 124006 (2016).

\bibitem{Bernar} RP Bernar, LCB Crispino, and A Higuchi, Phys. Rev. D 97, 085005 (2018).

\bibitem{higuchi1} A. Higuchi and S.S. Kouris, Class. Quant. Grav. 18, 4317 (2001).

\bibitem{Garidi124028} T. Garidi, E. Huguet, J. Renaud, Phys. Rev. D 67, 124028, (2003), and references herein.

\bibitem{Hooft} G. 't Hooft, arXiv:gr-qc/9310026; L. Susskind, J. Math. Phys. 36, 6377 (1995) [arXiv:hep-th/9409089].

\bibitem{Gibbons} G.W. Gibbons and S.W. Hawking, Phys. Rev. D 15, 2738 (1977).

\bibitem{Garidi2005} T. Garidi, E. Huguet, and J. Renaud, J. Phys. A 38, 245 (2005).

\bibitem{blackhole} H. Pejhan and S. Rahbardehghan, Int. Jour. Mod. Phys. A 31, 1650052 (2016).

\bibitem{braneII} H. Pejhan and S. Rahbardehghan, Phys. Rev. D 94, 064034 (2016).

\bibitem{braneI} S. Rahbardehghan and H. Pejhan, Phys. Lett. B 750, 627 (2015).

\bibitem{Casimir} H. Pejhan, M.R. Tanhayi and M.V. Takook, Ann. Phys. 341, 195 (2014).

\bibitem{WaldBook} R.M. Wald, \emph{Quantum Fields Theory in Curved Spacetime and Black Hole Thermodynamics}, The University of Chicago Press, 1994.

\end{thebibliography}
\end{document}